\documentclass[aps,prl,twocolumn,showpacs,groupedaddress,a4paper]{revtex4}  
\usepackage{graphicx}  
\usepackage{bm}        
\usepackage{amssymb}   

\newcommand{\subscript}[1]{\ensuremath{_{\textrm{#1}}}}

\begin{document}

\title{Optical manipulation of the wave function of quasiparticles in a solid}

\date{\today}

\author{R. Cerna$^{1*}$, D. Sarchi$^2$, T. K. Para\"{\i}so$^1$, G. Nardin$^1$, Y. L\'eger$^1$, M. Richard$^3$, B. Pietka$^1$, O. El Daif$^4$, F. Morier-Genoud$^1$, V. Savona$^2$, M. T. Portella-Oberli$^1$, B. Deveaud-Pl\'edran$^1$}

\affiliation{
$^1$Institut de Photonique et d'{\'E}lectronique Quantiques, {\'E}cole Polytechnique F{\'e}d{\'e}rale de Lausanne (EPFL), CH-1015 Lausanne; $^*$\textbf{roland.cerna@epfl.ch}\\
$^2$Institute of Theoretical Physics, Ecole Polytechnique F\'ed\'erale de Lausanne EPFL, CH-1015 Lausanne, Switzerland\\
$^3$Institut N\'eel-CNRS, 25 Avenue des Martyrs, BP 166, 38042 Grenoble Cedex 9, France\\
$^4$Institut des Nanotechnologies de Lyon (INL), UMR CNRS 5270, Ecole Centrale de Lyon, 36 Avenue Guy de Collongue, 69134 Ecully Cedex, France}


\maketitle 

{\bf Polaritons in semiconductor microcavities are hybrid quasiparticles consisting of a superposition of photons and excitons. Due to the photon component, polaritons are characterized by a quantum coherence length in the several micron range. Owing to their exciton content, they display sizeable interactions, both mutual and with other electronic degrees of freedom. These unique features have produced striking matter wave phenomena, such as Bose-Einstein condensation\cite{Kasprzak2006,Deng2007}, or parametric processes \cite{Stevenson2000,Savvidis2000} able to generate quantum entangled polariton states \cite{Savasta2005,Romanelli2007}. Recently, several paradigms for spatial confinement of polaritons in semiconductor devices have been established \cite{Kaitouni2006,Bajoni2007,Balili2007,Lai2007}. This opens the way to quantum devices in which polaritons can be used as a vector of quantum information  \cite{Quinteiro2006}. An essential element of each quantum device is the quantum state control. Here we demonstrate control of the wave function of confined polaritons, by means of tailored resonant optical excitation. By tuning the energy and momentum of the laser, we achieve precise control of the momentum pattern of the polariton wave function. A theoretical model supports unambiguously our observations.}

The key feature of polaritons in respect of confinement is their very small effective mass, which is about $10^5$ times smaller than the free electron mass. Hence, confinement over a few microns is enough to produce an atom-like spectrum with discrete energy levels. The photonic component of polaritons also allows control of the electronic excitations by optical means. These features hold great promise for devices where quantum correlations can be generated and controlled over long spatial range \cite{Quinteiro2006}.

 \begin{figure}
		\includegraphics[width=0.48\textwidth]{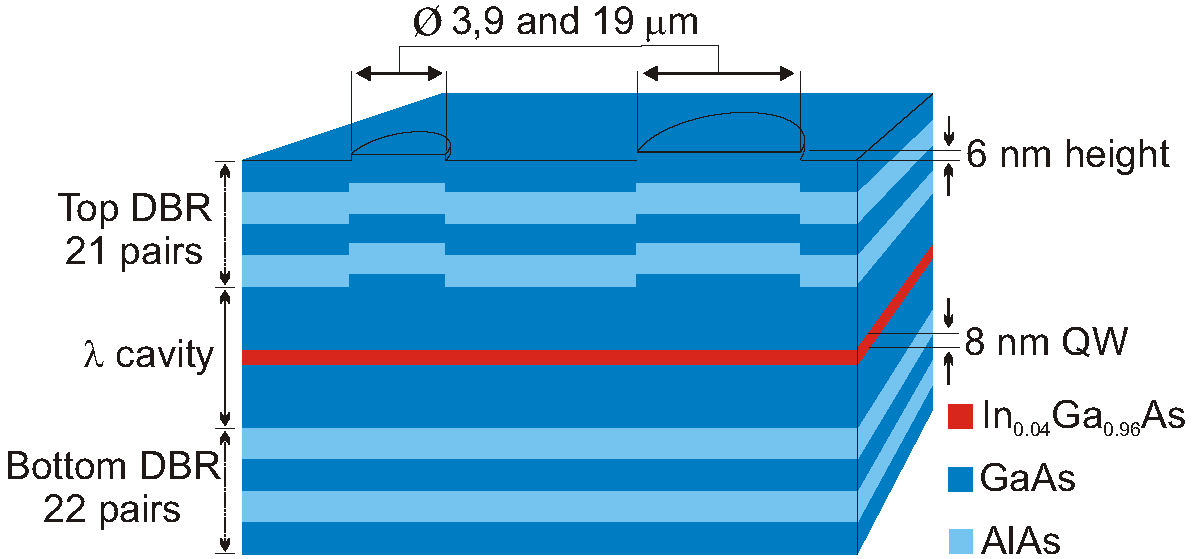}
	\caption{Sketch of the sample (not to scale). Single quantum well embedded in a $\lambda$ cavity with the exciton resonance at 1.484 eV. The sample has mesas of three different sizes and is wedge shaped.}
	\label{fig:sample1}
\end{figure}

The sample we used to carry out our studies consists of a single quantum well (QW) embedded in a planar microcavity (MC) with a mesa pattern on the spacer layer \cite{El2006} (see also Fig. 1 and Methods). Patterning the cavity thickness results in a modulation of the photon resonance energy, which corresponds to a potential trap, with finite energy barriers, able to confine polaritons. The confined polariton modes are then characterized by a discrete energy spectrum and spatially confined patterns \cite{Kaitouni2006,Baas2006}. We excited the sample optically with a cw laser and performed 2D k-space imaging and spatially resolved spectroscopy. All the studies we present here have been carried out in the linear response regime.

\begin{figure*}[h!,t]
		\includegraphics[width=0.5\textwidth]{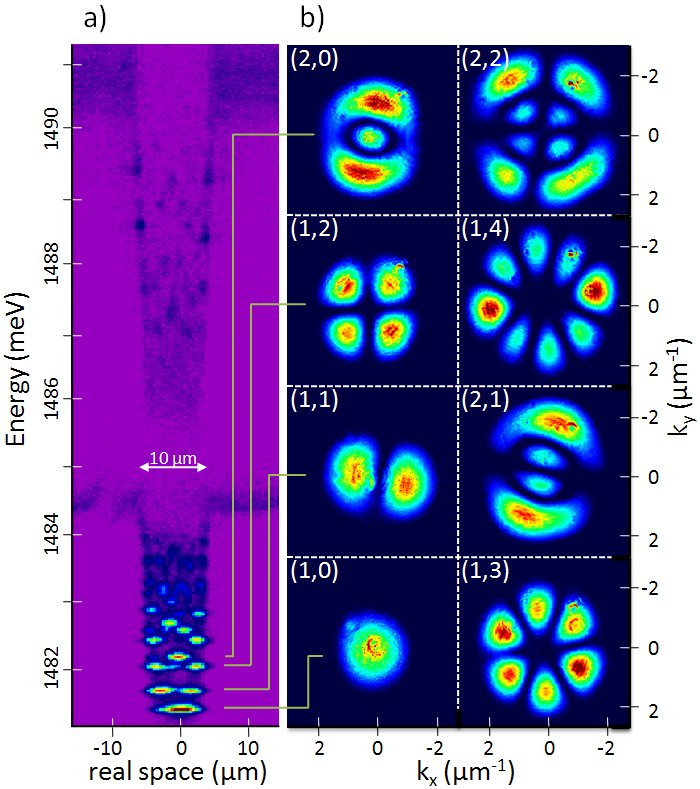}
	\caption{a) real space spectrum of a 9  $\mu$m mesa at 2 meV negative detuning (see Methods for detuning definition); upper and lower polariton branches of the 2D and 0D polaritons are clearly visible. b) k-space images of resonantly excited states from the 0D lower polariton branch. The states can be labeled with two quantum numbers $n$ and $m$. $2\left|m\right|$ gives the number of nodes along $\phi$ on a full circle, while $n$ corresponds to the number of lobes in radial direction.\label{fig:spec}}	
\end{figure*}

Figure \ref{fig:spec}a shows the spatially resolved photoluminescence spectrum of a 9 $\mu$m mesa at negative detuning, under non-resonant excitation. One can clearly distinguish the confined lower (1481.3-1484 meV) and upper (1485.5-1489.5 meV) polariton states in the mesa as well as the continuum states of the polaritons from the surrounding planar MC. This energy spectrum corresponds well to what is expected from theory for a 9 $\mu$m trap with cylindrical symmetry \cite{Kaitouni2006,Baas2006}. Figure \ref{fig:spec}b shows the emission patterns in the 2D k-space of different confined lower polariton states. For this kind of measurements we excited the states resonantly in energy and $\textbf{k}_\|$. We controlled the in-plane momentum $\textbf{k}_\|= (k_x,k_y)$ transferred to the polaritons by varying the excitation angle $\theta$ of the laser beam ($k_\| \propto \sin (\theta) / \lambda_{\textrm{\small{laser}}}$). The observed emission patterns are proportional to the momentum probability distribution of the excited states thus allowing us to access the squared absolute value $\left|\psi\right|^2$ of their wave functions. For resonant excitation, the emission is dominated by the resonant linear response, while the photoluminescence due to energy relaxation is negligible \cite{Hayes1998}.

The eigenstates $\left|n,m\right\rangle$ can be labeled by a radial quantum number $n$ $(1,2,3 \ldots)$ and an angular quantum number $m$ $(0, \pm1, \pm2 \ldots)$ with no restriction over $m$ \cite{Combescot2001}. The $\pm m$ states are degenerate due to the cylindrical symmetry of the potential. This symmetry should be visible in the observed emission pattern. Figure (\ref{fig:spec}b) shows that the emission at the energies of the states with $\left|m\right|>0$ and hence the polariton wave functions of these states feature lobes along the angular direction $\phi$. These lobes are evidence of a breaking of the symmetry, which can come either from an anisotropy of the confinement potential or from the incidence angle of the excitation beam. We will see that the anisotropy plays a role only for the $m=\pm1$ doublet of the 3 $\mu$m mesa. In 9 $\mu$m mesas the anisotropy of the potential can be neglected in first approximation. The specific pattern comes then from the coherent excitation of the $\pm m$ states at a $k_\| \neq 0$. The phases of the two eigenstates are locked at the laser position in k-space. Due to the angular dependence of the eigenstates wave functions $(\psi_{\pm m} \propto e^{\pm m i \phi})$, their phases evolve with opposite signs along $\phi$. Since the observed image is proportional to $\left|\psi_{+m} +  \psi_{-m}\right|^2$, the phase difference of the two components gives rise to positive and negative interferences along $\phi$. The number of nodes on a full circle is hence equal to $2\left|m\right|$. The number of lobes in radial direction corresponds to $n$. The same conclusions hold for all degenerate $m$-doublets in all investigated mesas.

\begin{figure*}
		\includegraphics[width=0.6\textwidth]{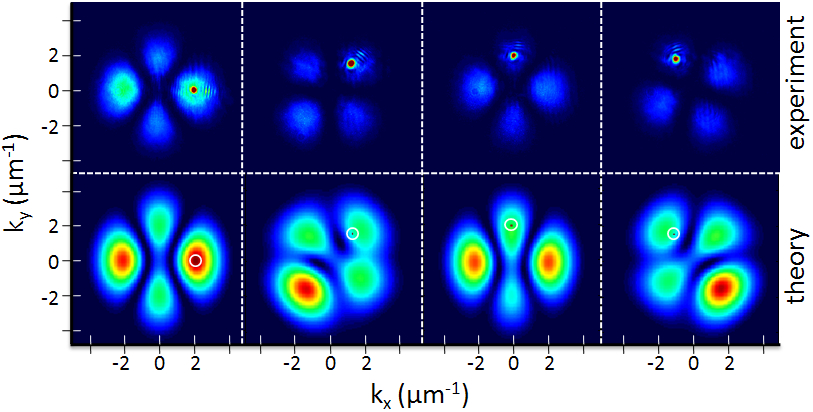}
	\caption{Lobes of the degenerate second doublet state in a 3 $\mu$m mesa following the rotation of the laser in k-space. The mesa was at zero detuning. The upper part shows the experimental results and the theoretical results from the simulation are shown in the lower part of the figure. The red spot which is visible in each experimental image of this figure is an emission maximum arising from the laser light transmitted directly through the cavity beside the mesa. This makes it easy to observe the excitation angle. In the simulation images the laser position is marked by a circle.\label{fig:4lobesrot}}
\end{figure*}

This phase locking effect can be used to manipulate the wave functions of degenerate doublet states. By changing the position of the excitation along $\phi$ in k-space, the phase of the $\pm m$ components is changed, and thus the interference pattern. This is demonstrated in Fig. \ref{fig:4lobesrot} for the second excited state $(1,\pm2)$ in a 3 $\mu$m mesa.  The intensity lobes follow the rotation of the laser. This demonstrates control over the probability distribution of polaritons in momentum space and also in real space. 

Unlike the larger mesas, the 3 $\mu$m mesas have a non negligible ellipticity $(b/a \approx 1.29)$, which can lift the degeneracy of $m=\pm 1$ doublet states. This is the case for the first eigenstate doublet $\left|a\right\rangle$ and $\left|b\right\rangle$, which can be pictured as linear combinations of the $m=\pm 1$ states of a cylindrical structure with the same average diameter: $\left|a\right\rangle=\frac{1}{\sqrt{2}}(\left|1,-1\right\rangle+\left|1,+1\right\rangle)$ and $\left|b\right\rangle=\frac{1}{\sqrt{2}}(\left|1,-1\right\rangle-\left|1,+1\right\rangle)$. Therefore these eigenstates feature lobes along $\phi$ and their orientation is fixed by the axis of the ellipse. A degeneracy lift has only been observed for this lowest doublet state, as expected from theory.
The splitting of the non-degenerate doublet increases with negative detuning due to the decreasing effective mass and is about $0.11\pm 0.04$ meV at zero detuning. This agrees with the 0.08 meV expected from our model. If the splitting is large enough ($>0.1$ meV) one can excite independently the $\left|a\right\rangle$ or $\left|  b\right\rangle$ state. In this case the lobes do not follow the rotation of the laser since there is no longer interference between two eigenstates. The lobes will just vary in intensity depending on the overlap between the plane wave of the laser and the wave function of the eigenstate. 

\begin{figure*}[h!,t]
		\includegraphics[width=0.6\textwidth]{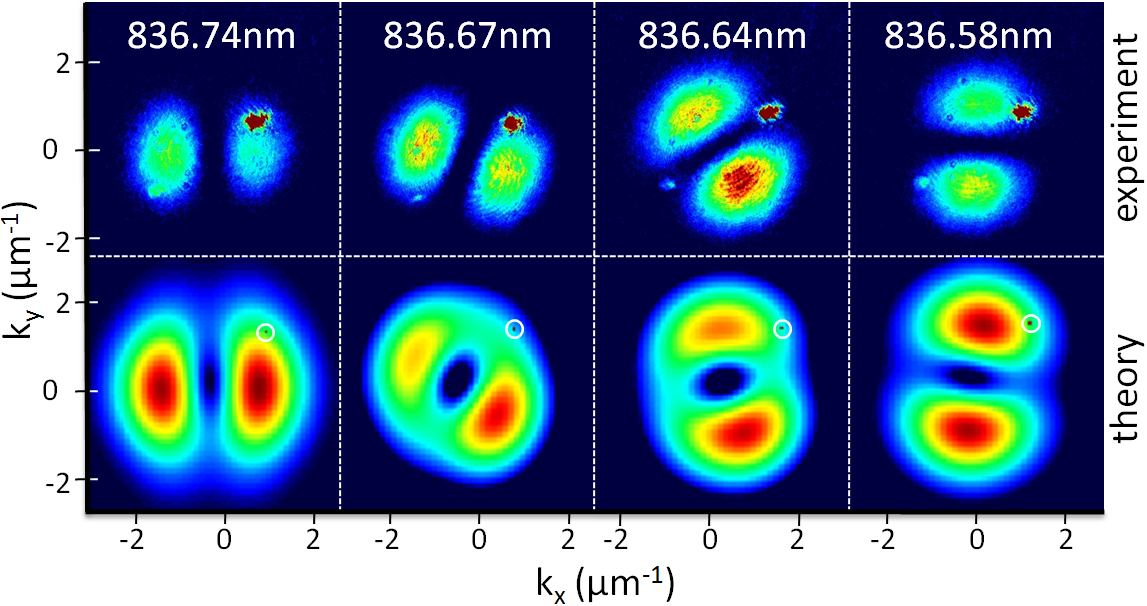}
	\caption{Manipulation of the wave functions of the split first doublet in a 3 $\mu$m mesa ($\approx 3$ meV negative detuning) by changing the excitation energy while keeping the excitation point constant. Upper part: experiment; Lower part: theoretical results. The polariton lifetime in our sample is about 15 ps, which leads to an energy linewidth of the eigenstates much broader than the spectral width of the laser. The lobes rotate due to the different contributions of the eigenstates to the linear response. \label{fig:energymanip}}
\end{figure*}

Depending on the detuning, the momentum distribution of these split states can nevertheless be manipulated. At negative detuning the energy splitting can be significantly larger than the natural linewidth of the modes ($\approx$ 85 $\mu$eV). As the laser spectral width is smaller ($<$ 25 $\mu$eV), it is possible to excite only one state, but with different energies. Then the measured momentum pattern is that of the corresponding eigenstate. In the case of a split doublet with sizeable spectral overlap, the laser energy can be tuned between the eigenenergies of the two split states. As a consequence, it is then possible to vary the relative amplitude and phase of the two contributions to the linear response thanks to the different overlap of the laser with the two states. In this case one can rotate the polariton momentum distribution by rotating the laser momentum $\textbf{k}_\|$ as shown before and also by varying the laser energy. This offers additional control: by tuning the laser energy one can control the orientation of the lobes with respect to the laser position in k-space. Figure \ref{fig:energymanip} shows the results of an experiment where we kept the excitation position in k-space constant and changed the excitation energy between two split states with significant spectral overlap.

As the figures \ref{fig:4lobesrot} and \ref{fig:energymanip} show, the experimental results presented here can be reproduced with a theoretical model which is described into more detail in \cite{sarchi09}. We start from the bosonic exciton-photon Hamiltonian \cite{sarchi07,ben01} and we adopt a linearized mean-field approach~\cite{carusottociuti05}.
In this approach and considering a coherent monochromatic optical pump $F({\bf r},t)=e^{-i\omega t}F_0({\bf r})$, the coherent exciton and photon fields $\Phi_{x,c}({\bf r},t)$ evolve with the frequency of the laser source as $\Phi_{x,c}({\bf r},t)=e^{-i\omega t}\Phi_{x,c}^0({\bf r})$. Their steady-state shape $\Phi_{x,c}^0({\bf r})$ obeys a set of two coupled Schr\"odinger-like equations for the exciton-photon problem \cite{sarchi07,carusottociuti05}
\begin{equation}
\left(-\frac{\hbar^2\nabla^2}{2 m_x}-\hbar\omega-i\gamma_x\right)\Phi^0_x({\bf r})=\hbar\Omega_R\Phi^0_c({\bf r})\,, \label{eq:GP0_X}
\end{equation}
\begin{equation}
\left(\epsilon_c(\vec{\nabla})+U_{c}({\bf r})-\hbar\omega-i\gamma_c\right)\Phi^0_c({\bf r})=\hbar\Omega_R\Phi^0_x({\bf r})+F({\bf r})\,, \label{eq:GP0_C}
\end{equation}
where $m_x$ is the exciton mass, $\epsilon_c(k)$ is the cavity photon dispersion accounting for the photon-exciton detuning, $U_c$ is the photon confining potential,
$\hbar\Omega_{R}$ is the exciton-photon coupling and the decay rates $\gamma_{x,c}$ account for the finite lifetime of the two species. Here we are omitting the non-linear terms because we are interested in the linear regime (see discussion below).
From these equations, it is clear that the phase of the coherent fields $\Phi_{x,c}({\bf r},t)$ generated by the coherent pumping is locked to the phase of the laser source. The coherent fields in k-space are obtained from the fields $\Phi_{x,c}^0({\bf r})$ via Fourier transform.

From Eqs. (\ref{eq:GP0_X}) and (\ref{eq:GP0_C}), we see that the exciton coherent field is expected to have the same shape as the photon field, i.e. $\Phi_x^0({\bf r})\equiv X \Phi_p^0({\bf r})$ and $\Phi_c^0({\bf r})\equiv C \Phi_p^0({\bf r})$, as a result of the exciton-photon coupling. The relative intensity defines the exciton amount as
\begin{equation}
\mid X\mid^2=\frac{\int d{\bf r}\mid\Phi_{x}^0({\bf r})\mid^2}{\int d{\bf r}\mid\Phi_{p}^0({\bf r})\mid^2} \,,
\end{equation}
where $|\Phi_{p}^0({\bf r})|^2=\mid\Phi_{x}^0({\bf r})\mid^2+\mid\Phi_{c}^0({\bf r})\mid^2$. This clearly proves that, since $|X|^2\sim 0.5$, the reported manipulation of the emitted optical field also implies the manipulation of the matter field.

For the calculations, we use a confining potential in agreement with AFM images of the mesas, thus accounting for the weak ellipticity of the samples. The decay rates are assumed to be $\gamma_c=0.05~\mbox{meV}$, in agreement with the reported photon lifetime $\tau_c=15~\mbox{ps}$, and $\gamma_x=0.01~\mbox{meV}$. The other parameters are obtained from the experimental characterization of the sample.

The effect of polariton-polariton interactions could be very important, especially in presence of spatial confinement. By including the non-linear terms in Eqs. (\ref{eq:GP0_X},\ref{eq:GP0_C}) \cite{carusottociuti05} we have checked that this is not the case for the pump intensities used in this experiment. Therefore, all the reported features are entirely due to the linear response of the polariton system to resonant laser excitation.

\section{Methods}
\footnotesize
 The investigated sample consists of a single semiconductor quantum well (QW) placed in the middle of a $\lambda$ microcavity (MC) (see Fig. \ref{fig:sample1}). The QW, which is composed of a 8 nm thick In$_{0.04}$Ga$_{0.96}$As layer sandwiched between two GaAs barrier layers, shows a sharp exciton resonance (500 $\mu$eV FWHM) at E\subscript{x} = 1.4846 eV.  The two distributed Bragg reflectors (DBRs) of the MC are made of 21 (top) and 22 (bottom) layer pairs of AlAs/GaAs.  The resonance energy of the cavity $\textrm{E}_c = ch/\lambda_c$ is given by the cavity spacer thickness $\lambda_c/n$. The sample is wedge-shaped thus allowing to change the energy detuning $(\delta = \textrm{E}_c - \textrm{E}_X)$ between the cavity photons and the QW excitons as a function of sampling position. The Rabi splitting (the energy difference between the upper and lower polariton branch at $\textbf{k}_\|=0$) has a value of 3.5 meV at zero detuning. Round mesas of different sizes (3, 9 and 19 $\mu$m in diameter) have been etched on the top of the spacer before growing the upper DBR \cite{El2006}. This leads to a locally lowered resonance energy $\textrm{E}_c =\hbar \omega_c$ of the cavity. The mesas have a height of 6 nm, which corresponds to a potential depth of about 9 meV at the exciton energy \cite{Kaitouni2006}. Thus the mesas act as a three dimensional trap for MC photons. The confined photon modes couple to the 2D excitons thus creating the upper and lower confined polariton branches \cite{Panzarini1999}. The anti-crossing behaviour of all confined polariton modes has been shown in \cite{Baas2006}. When talking about detuning in the context of confined polaritons we always refer to the energy difference between the lowest confined photon mode and the 2D excitons. At zero detuning for example the ground state is half light half matter while the excited states of the confined lower polaritons have an higher excitonic component.  
 
To carry out our study we employed a photoluminescence setup in transmission configuration. The polaritons have been optically excited with a tunable cw Ti:Sa laser. For non resonant excitation the laser has been tuned at the first reflectivity minimum of the DBR's around 784 nm. The laser beam has been focused on the sample with a camera objective (f=55 mm and 1.2 aperture) providing an excitation spot of about 25 to 30 $\mu$m for all measurements. This provides a narrow distribution in k-space. The excitation power, which was about 5 mW for non resonant excitation and about 7 $\mu$W for resonant excitation, was chosen in such a way that the system would always be in the linear regime. By displacing the excitation beam with a retroreflector parallel to the middle axis of the camera lens we changed the incidence angle and thus the in-plane momentum $\textbf{k}_\|= (k_x,k_y)$ transferred to the polaritons.

The sample was kept in a continuous flow optical cryostat at around 5 K. On the detection side the photoluminescence light was collected with a microscope objective with high numerical aperture (NA = 0.55). The collimated luminescence light was split by a beam splitter. One part of the beam was dispersed by a spectrometer (25 $\mu$eV resolution) and recorded with a nitrogen cooled CCD camera. The second part of the beam was sent to an angle resolved imaging setup in order to measure the in-plane momentum $\textbf{k}_\|= (k_x,k_y)$ of the resonantly excited polaritons. The emission angle of the luminescence light from a MC is proportional to the in plane momentum of the emitter (polariton) $k_\| \propto \sin (\theta) / \lambda_c$.

\section{Acknowledgements} This work was supported by the Quantum Photonics NCCR and the Swiss National Science
Foundation.


\begin{thebibliography}{99}

\bibitem{Kasprzak2006} Kasprzak, J. {\sl et al.} Bose-Einstein condensation of exciton polaritons \textit{Nature} 443, 409 (2006).
\bibitem{Deng2007} Deng, H., Solomon, G.S., Hey, R., Ploog, K.H. \& Yamamoto, Y. Spatial coherence of a polariton condensate \textit{Phys. Rev. Lett.} 99, 126403 (2007).
\bibitem{Stevenson2000} Stevenson, R.M.  {\sl et al.} Continuous wave observation of massive polariton redistribution by stimulated scattering in semiconductor microcavities \textit{Phys. Rev. Lett.} 85, 3680 (2000).
\bibitem{Savvidis2000} Savvidis, P.G. {\sl et al.} Angle-Resonant Stimulated Polariton Amplifier \textit{Phys. Rev. Lett.} 84, 1547 (2000).
\bibitem{Romanelli2007} Romanelli, M., Leyder, C., Karr, J.Ph., Giacobino, E. \& Bramati, A. Four wave mixing oscillation in a semiconductor microcavity: Generation of two correlated polariton populations \textit{Phys. Rev. Lett.} 98, 106401 (2007).
\bibitem{Savasta2005} Savasta, S., Di Stefano, O., Savona, V. \& Langbein, W. Quantum complementarity of microcavity polaritons \textit{Phys. Rev. Lett.} 94, 1 (2005).
\bibitem{Bajoni2007} Bajoni, D. {\sl et al.} Polariton parametric luminescence in a single micropillar \textit{Appl. Phys. Lett.} 90, 051107 (2007).
\bibitem{Balili2007} Balili, R., Hartwell, V., Snoke, D., Pfeiffer, L. \& West, K. Bose-Einstein condensation of microcavity polaritons in a trap \textit{Science} 316, 1007 (2007).
\bibitem{Lai2007} Lai, C. W. {\sl et al.} Coherent zero-state and $\pi$-state in an exciton-polariton condensate array \textit{Nature} 450, 529 (2007).
\bibitem{Kaitouni2006} Kaitouni, R.I. {\sl et al.} Engineering the spatial confinement of exciton polaritons in semiconductors  \textit{Phys. Rev. B}, 74, 15 (2006). 
\bibitem{El2006} El Daif, O. {\sl et al.} Polariton quantum boxes in semiconductor microcavities \textit{Appl. Phys. Lett.} 88, 061105 (2006).
\bibitem{Quinteiro2006} Quinteiro, G.F., Fern\'andez-Rossier, J., \& Piermarocchi, C. Long-range spin-qubit interaction mediated by microcavity polaritons \textit{Phys. Rev. Lett.} 97, 097401 (2006).
\bibitem{Baas2006} Baas, A. {\sl et al.} Zero dimensional exciton-polaritons \textit{Phys. Status Solidi B}, 243, 2311 (2006).
\bibitem{Hayes1998} Hayes, G.R. {\sl et al.} Resonant Rayleigh scattering versus incoherent luminescence in semiconductor microcavities \textit{Phys. Rev. B}, 58, R10175 (1998).
\bibitem{Combescot2001} Leyronas, X. \& Combescot, M. Quantum wells, wires and dots with finite barrier: Analytical expressions for the bound states \textit{Solid State Commun.}, 119, 631--635 (2001).
\bibitem{sarchi09} Sarchi, D. {\sl et al.} Polariton parametric photoluminescence in spatially inhomogeneous systems \textit{Phys. Rev. B}, 79, 165315 (2009).
\bibitem{sarchi07} Sarchi, D. \& Savona, V. Spectrum and thermal fluctuations of a microcavity polariton Bose-Einstein condensate \textit{Phys. Rev. B}, 77, 045304 (2008).
\bibitem{carusottociuti05} Carusotto, I. \& Ciuti, C. Spontaneous microcavity-polariton coherence across the parametric threshold: Quantum Monte Carlo studies \textit{Phys. Rev. B}, 72, 125335 (2005).
\bibitem{ben01} Ben-Tabou de-Leon, S. \& Laikhtman, B. Exciton-exciton interactions in quantum wells: Optical properties and energy and spin relaxation \textit{Phys. Rev. B}, 63, 125306 (2001).
\bibitem{Panzarini1999} Panzarini, G. \& Andreani, L.C. Quantum theory of exciton polaritons in cylindrical semiconductor microcavities \textit{Phys. Rev. B}, 60, 16799 (1999).


\end{thebibliography}
\end{document}